\documentclass[final,4p,11pt,onecolumn]{elsarticle}
\usepackage{setspace}
\usepackage{cite}
\usepackage{graphicx}
\usepackage{subfigure}
\usepackage{amsmath}
\usepackage{amssymb}
\usepackage{amsthm}
\usepackage{algorithm}
\usepackage{booktabs}

\newcommand{\bdsb}[1]{\boldsymbol{#1}}

\DeclareMathOperator*{\argmax}{arg\,max}

\newtheorem{thm}{Theorem}

\newtheorem{corr}{Corrolary}

\title{\Huge GPS Signal Acquisition via Compressive Multichannel Sampling}

\author{Xiao Li$^{\dagger}$,~Andrea Rueetschi$^{\dagger}$, Yonina C. Eldar$^{\star}$ and Anna Scaglione$^{\dagger}$}

\address{$^{\dagger}$ University of California, Davis, email : \{eceli,arueetschi,ascaglione\}@ucdavis.edu\\
$^{\ast}$ Technion - Israel Institute of Technology, Haifa, email : yonina@ee.technion.ac.il}

\begin{document}
\graphicspath{{./sim_plots/}}

\begin{frontmatter}
%\date{}
%\maketitle

\begin{abstract}
In this paper, we propose an efficient acquisition scheme for GPS receivers. It is shown that GPS signals can be effectively sampled and detected using a bank of randomized correlators with much fewer chip-matched filters than those used in existing GPS signal acquisition algorithms. The latter use correlations with all possible shifted replicas of the satellite-specific C/A code and an exhaustive search for peaking signals over the delay-Doppler space. Our scheme is based on the recently proposed analog compressed sensing framework, and consists of a multichannel sampling structure with far fewer correlators.

The compressive multichannel sampler outputs are linear combinations of a vector whose support tends to be sparse; by detecting its support one can identify the strongest satellite signals in the field of view and pinpoint the correct code-phase and Doppler shifts for finer resolution during tracking. The analysis in this paper demonstrates that GPS signals can be detected and acquired via the proposed structure at a lower cost in terms of number of correlations that need to be computed in the coarse acquisition phase, which in current GPS technology scales like the product of the number of all possible delays and Doppler shifts. In contrast, the required number of correlators in our compressive multichannel scheme scales as the number of satellites in the field of view of the device times the logarithm of number of  delay-Doppler bins explored, as is typical for compressed sensing methods.
\end{abstract}

\begin{keyword}
GPS, compressive sensing, spread spectrum, analog compressed sensing
\end{keyword}

\end{frontmatter}

\newpage
\section{Introduction}
Nowadays, Global Positioning System (GPS) chips are ubiquitous, and continue to be embedded in a variety of devices. A GPS device allows to determine its location with about $3$ meters accuracy, by measuring the propagation delay of signals transmitted by the set of GPS satellites in the field of view (FOV) of any receiver located on the surface of the earth, which typically requires measurements from at least four satellites \citep{Kaplan}.

Conventionally, the signal that arrives at the receiver is downconverted, match-filtered and oversampled at a fast rate. Subsequently, the receiver acquires enough (at least four) strong signals by exploiting the orthogonality of the distinct {\it coarse/acquisition} (C/A) codes used in GPS signaling at each satellite \citep{Braasch}. However, due to the unknown propagation delays, the samples obtained are misaligned in time and frequency and therefore, it is vital to pinpoint the code-phase in order to decode the navigation data correctly \citep{Braasch}\citep{Auber} and use the time-delay information for pseudo-range computation. Furthermore, each of the satellites contributes a  component of the received GPS signal that is characterized by a distinct Doppler offset \citep{Ohmori}, due to the unequal relative velocity of satellite and receiver, as well as the offset of the different local oscillators at the GPS receivers. In general,  time-frequency synchronization as well as signal detection is tackled in GPS receivers during the {\it acquisition/detection} stage via a parallel search over the binned delay-Doppler space across all the satellite C/A codes \citep{Ward}\citep{Sahmoudi}.

In many practical scenarios, signals might arrive at the receiver with multipath components instead of the line of sight (LOS) component \citep{Auber}\citep{Sahmoudi}\citep{Soubielle}. Constructive and destructive superposition of randomly delayed and faded replicas, leads to distorted correlation peaks. This is usually tackled in the {\it tracking stage} \citep{Braasch} that follows the {\it acquisition/detection} stage, by using an {\it early-late} receiver. Such a receiver compares the energy of a symbol period in the first half from the early gate to the energy in the last half from the late gate so that the receiver can synchronize the signals accordingly. Furthermore, many approaches, in addition to the early-late structure, have been proposed to better mitigate the effects brought by multipath, including (but not limited to) the Narrow Correlator \citep{Enge}, Multipath Eliminating Technique (MET) \citep{Townsend}, and Multipath Estimating Delay Lock Loop (MEDLL) \citep{Nee}. These methods differ in their capabilities to remove multipath errors, specifically at low signal-to-noise ratio (SNR) and/or in the presence of interference. In this work, we consider the general signal model that considers multipath effects and propose an acquisition scheme that coarsely captures significant paths for each active satellite, with its corresponding code-phase and Doppler. The tracking stage that further resolves the estimates of delay-Doppler pairs as well as the multipath components is beyond the scope of this paper.

As described above, the acquisition and detection of GPS signals is usually performed sequentially. First, the strongest signals coming from the satellites are detected by searching a binned delay-Doppler space via exhaustive correlations that pinpoint the correct coarse timing information and frequency offsets. After acquisition and detection, the signal is locked and the device enters the tracking stage that tackles fine synchronization and multipath error mitigation in order to despread, demodulate and decode the navigation data correctly in real-time. However, this acquisition/detection scheme can be computationally intensive due to the large number of correlations, and especially the exhaustive search for peaks over the binned delay-Doppler space across all the satellite signals with distinct C/A codes. For example, the maximum Doppler shift in a GPS signal is typically within $[-10{\rm kHz},10{\rm kHz}]$ and the search step size is usually $500$Hz while the maximum delay can run up to a C/A code length $1023$. In this case, the 2-D delay-Doppler peak is found by comparing the outputs of $1023 \times 41\approx 4\times 10^4$ correlators for each satellite, which is a heavy computation burden.

{\it Paper contributions:} In order to scale down the operations and hardware requirements, we propose a simple and efficient acquisition scheme based on the recently developed compressed sensing (CS) framework \citep{Tao} and its extension to analog signals \citep{Eldar_CS}. The multichannel samplers in \citep{Eldar_CS} are constructed as a randomized linear combination of the duals of all the generators, where the generators in this case correspond to the satellite-specfic C/A code waveforms. In our context, we show that the the duals of the generators are well approximated by the generators themselves. This alleviates one of the most difficult aspects in the practical application of \citep{Eldar_CS}, namely, the physical implementation of the dual filters, by exploiting properties of the spread spectrum sequences that are in the GPS standard. Thanks to this interpretation, the proposed multichannel samplers can be viewed as performing independent random projections of all correlators outputs. The resulting set of compressive measurements are then used together to recover the peaks located sparsely over the delay-Doppler space, which is a jointly sparse recovery problem with infinite input vectors and infinite measurement vectors (IMV). The continuous-to-finite (CTF) method introduced in \citep{Eldar_MMV} effectively reduces the IMV problem to a finite multiple measurement vector (MMV) system with jointly sparse inputs, which can be solved efficiently using the Reduce MMV and Boost (ReMBo) technique proposed in \citep{Eldar_MMV}, or other MMV approaches \citep{Cotter}\citep{Chen}.

The paper is organized as follows. Section 2 describes the general model for GPS signals. Section 3 re-interprets existing GPS acquisition schemes from a sampling point of view. In order to scale down the computations and hardware requirements, Section 4 introduces the analog CS framework. In Section 5 we further reduce the general solution to a set of simple compressive samplers by utilizing the structure of GPS signals. Numerical results are shown in Section 6 to demonstrate the effectiveness of our proposed acquisition scheme, followed by a complexity analysis given in Section 7. Finally the paper is concluded in Section 8.

\section{GPS Signal Model}\label{signal_model}
The signal transmitted by the satellites is a direct sequence spread spectrum (DS-SS) signal modulated onto $L1$ and $L2$ frequencies at $1575.42$MHz and $1227.60$MHz respectively. In commercial GPS systems publicly available to civil users, the DS-SS signal received at the user end is carried on $L1$ frequency from all the available launched satellites. Equivalently, the baseband signal from the $i$th satellite is transmitted as
\begin{align}\label{sig_model}
	s_i(t) = \sum_{n\in\mathbb{Z}}d_i[n]\phi_i(t-nT),\quad i = 1,\cdots,I
\end{align}
where $\phi_i(t)$ is a spreading waveform determined by a satellite-specific spreading code and  $\{d_i[n]\}_{n\in\mathbb{Z}}$ is the navigation data sent by the $i$th satellite with a symbol period of $T$, containing its time stamp, orbit location and relevant information entailed for positioning the receiver.

More specifically, the waveform $\phi_i(t)$ is determined by the $i$th satellite's C/A code $\{s_i[m]\}$ as
\begin{align}
	\phi_i(t) = \sum_{m=0}^{M-1}s_i[m]g(t-mT_c),\quad i = 1,\cdots,I
\end{align}
where $g(t)$ is a wideband short pulse. For simplicity, we assume that $g(t)$ has a flat spectrum of bandwidth $\Omega_g=2\pi L/T_c$ (typically $L=1$) approximated with error $\epsilon_g(\omega)$
\begin{align}\label{approx_spectrum}
	G(\omega)=\left[1+\epsilon_g(\omega)\right]\mathrm{rect}_{2\pi L/T_c}(\omega),
\end{align}
where $\epsilon_g(\omega)$ specifies the deviation from the flat spectrum with\footnote{The notation $\|\cdot\|$ refers to the $L_2$ norm of a function $\|\epsilon_g(\omega)\|\triangleq \sqrt{\int_{-\infty}^{\infty} |\epsilon_g(\omega)|^2\mathrm{d}\omega}$.} $\|\epsilon_g(\omega)\|\ll 1$. Due to the periodicity of the C/A code, $T=MT_c$. The C/A code $\{s_i[m]\}$ is a pseudo-random binary sequence of length $M$ that contains $N$ maximum length sequence (MLS) or Gold sequence of length $M_0=1023$ transmitted with a chip period $T_c=977.5$ns, which implies $M=NM_0$. In fact, by the GPS transmission standards we have $N=20$ for the GPS $L1$-C/A signal, i.e. $T=20$ms.

The correlation properties of the spreading code are vital in the recovery of spread spectrum signals. Denote the cross-correlation between different C/A code as
\begin{align}\label{R_ii}
    R_{i'i}[u]\triangleq \frac{1}{M}\sum_{m=0}^{M-1}s_{i'}[m-u]s_i^\ast[m].
\end{align}
When $M$ is large, the Gold sequences or MLS sequences are orthogonal between different satellites and approximately orthogonal between different shifts \citep{Soubielle}. This is indicated by the flat and $2\pi/T_c$-periodic {\it cross spectral density}
\begin{align}
	S_{i'i}\left(e^{\mathrm{i}\omega T_c}\right)
	\triangleq \sum_{u=-M+1}^{M-1}R_{i'i}[u]e^{-\mathrm{i} u \omega T_c}
	= \delta[i'-i]+\epsilon_{i',i}(\omega),
\end{align}
where the error function $\epsilon_{i',i}(\omega)$ is also $2\pi/T_c$-periodic with\footnote{The norm here is defined as $\|\epsilon_{i',i}(\omega)\|\triangleq\int_{-\pi/T_c}^{\pi/T_c}|\epsilon_{i',i}(\omega)|^2\mathrm{d}\omega $ due to the periodicity.} $\|\epsilon_{i',i}(\omega)\|\ll 1$. This flat property plays an essential role in simplifying the design presented later in this paper.

After downconversion, the signal at the receiver can be modeled as
\begin{align}
	x(t) = \sum_{i=1}^{I} \sum_{r=1}^{R} h_{i,r}s_i(t-\tau_{i,r}) e^{\mathrm{i} \omega_{i,r}t} +  v(t),
\end{align}
where $\{h_{i,r}\}_{r=1,\cdots,R}$ are the multipath channel taps with delays $\{\tau_{i,r}\}_{r=1,\cdots,R}$ and Doppler shifts $\{\omega_{i,r}\}_{r=1,\cdots,R}$ from the $i$th satellite to the receiver, and $v(t)$ is the Additive White Gaussian Noise (AWGN) with variance $\sigma^2$. Combined with the signal model (\ref{sig_model}), the signal $x(t)$ is represented as
\begin{align}
	x(t)
	= \sum_{n\in\mathbb{Z}} \sum_{i=1}^{I}
	     \sum_{r=1}^{R}a_{i,r}[n]\phi_i(t-nT-\tau_{i,r})e^{\mathrm{i} \omega_{i,r}t}+v(t),\nonumber
\end{align}
where $a_{i,r}[n]\triangleq h_{i,r}d_i[n]$. In the coarse acquisition phase, it is typically assumed that the delays are integer multiples of the chip duration $\tau_{i,r}=q_{i,r}T_c$ with $q_{i,r}\in\mathcal{Q}$ and the Doppler shifts are integer multiples of the frequency search step $\omega_{i,r}=k_{i,r}\Delta\omega$ with $k_{i,r}\in\mathcal{K}$, where the sets $\mathcal{Q}$ and $\mathcal{K}$ define the delay-Doppler space. This leads to the following discretized signal model
\begin{align}\label{discretized_model}
	x(t)	= \sum_{n\in\mathbb{Z}} \sum_{i=1}^{I}
	     \sum_{r=1}^{R}a_{i,r}[n]\phi_i(t-nT-q_{i,r}T_c)e^{\mathrm{i}k_{i,r}\Delta\omega t} +v(t).
\end{align}

\section{Standard GPS Acquisition Scheme}
The main task of the acquisition stage is to detect the correct code-phase $\bdsb{q}\triangleq\{q_{i,r}\}_{i=1,\cdots,I}^{r=1,\cdots,R}$ and Doppler shift $\bdsb{k}\triangleq\{k_{i,r}\}_{i=1,\cdots,I}^{r=1,\cdots,R}$ across the delay-Doppler space and recover the sequence $\{a_{i,r}[n]\}_{i=1,\cdots,I}$, among which the strongest set
$\mathcal{I}$ of satellites ($|\mathcal{I}|\geq 4$) are picked for the purpose of triangulation \citep{Kaplan} - \citep{Ohmori}. Note that the sequence $\{a_{i,r}[n]\}_{n\in\mathbb{Z}}$ includes the attenuation of the channels between the satellites and the receiver. Therefore, its magnitude indicates the strength of the signal received and only the strong ones are acquired by the receiver. In general, the magnitudes of those acquired $i\in\mathcal{I}$ are significantly greater than those $i\notin \mathcal{I}$, making the coefficients $a_{i,r}[n]$ sparse due to the wide difference in signal strength.

\subsection{Exhaustive Search via Matched Filtering (MF)}
Conventionally, the acquisition and detection of strong satellite signals is achieved by correlating the incoming signal $x(t)$ with a bank of match-filters $\phi_i(t)$'s that are separately modulated by carriers $\{e^{\mathrm{i} k \Delta\omega t}\}_{k\in\mathcal{K}}$ and shifted in time $\{\phi_i(t-qT_c)\}_{q\in\mathcal{Q}}$. In this way, the paths corresponding to peaks in the magnitude of $a_{i,r}[n]$ can be found in the delay-Doppler binned-space $\mathcal{Q}\times\mathcal{K}$ for each satellite corresponding to its C/A code.

This approach can be viewed as sampling with a set of filters, followed by uniform sampling at time $t=nT$, as depicted in Fig. 1. The sampling kernels of this equivalent structure are given by $\phi_{i,k,q}(t) =\phi_i(t-qT_c)e^{\mathrm{i}k\Delta\omega t}$,
for all $i=1,\cdots,I$, $k\in\mathcal{K}$ and $q\in\mathcal{Q}$. The sampled output in each channel is equal to
\begin{align}\label{z}
	z_{i,k,q}[n]\triangleq \langle x(t), \phi_{i,k,q}(t-nT) \rangle.
\end{align}
In the Fourier domain, we have
\begin{align}\label{C_DTFT}
	Z_{i,k,q}\left(e^{\mathrm{i}\omega T}\right) = \frac{1}{T}\sum_{\ell\in \mathbb{Z}}\Phi_{i,k,q}^\ast\left(\omega-\frac{2\pi \ell}{T}\right) X\left(\omega-\frac{2\pi\ell}{T}\right),
\end{align}
where $\Phi_{i,k,q}^\ast(\omega)$ and $X(\omega)$ are the Fourier transforms of $\phi_{i,k,q}(-t)$ and $x(t)$ respectively.
\begin{figure}
\center
\includegraphics[width=3in]{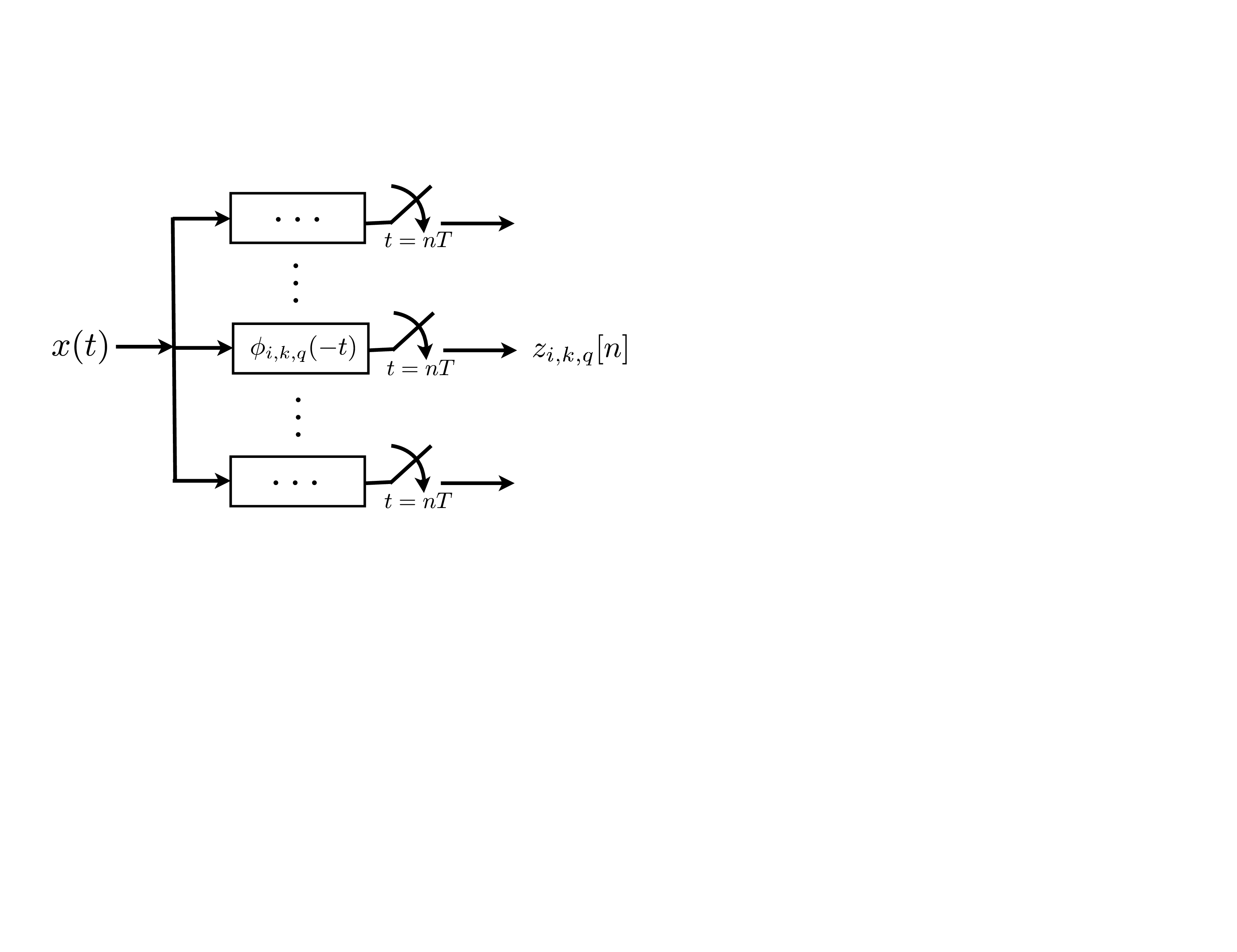}
\caption{Exhaustive Matched Filtering (MF) Approach}
\end{figure}
Note that the summation over $\ell\in\mathbb{Z}$ in (\ref{C_DTFT}) depends on the bandwidth of the filter $\phi_{i,k,q}(-t)$, where as mentioned in Section 2 the bandwidth of $g(t)$ is $\Omega_g=2\pi LM/T$. Therefore, the summation becomes finite from $\ell=0$ to $\ell=LM-1$ over $\omega\in[-\pi /T,\pi /T]$. From (\ref{discretized_model}), we can express $X(\omega)$ as
\begin{align}\label{x_Fourier}
	X(\omega) &= \sum_{i=1}^{I}\sum_{r=1}^{R}A_{i,r}\left(e^{\mathrm{i}\omega T}\right)\Phi_i(\omega-k_{i,r}\Delta\omega) e^{-\mathrm{i}(\omega-k_{i,r}\Delta\omega) q_{i,r}T_c}+V\left(e^{\mathrm{i}\omega T}\right),
\end{align}
where we defined $A_{i,r}\left(e^{\mathrm{i}\omega T}\right)\triangleq\sum_{n\in\mathbb{Z}}a_{i,r}[n]e^{-\mathrm{i}n(\omega-k_{i,r}\Delta\omega)T}$.
Substituting \eqref{x_Fourier} into \eqref{C_DTFT}, and denoting by $\bdsb{z}\left(e^{\mathrm{i}\omega T}\right)$ the length-$I|\mathcal{K}||\mathcal{Q}|$ column vector whose $(i,k,q)$th element is $Z_{i,k,q}\left(e^{\mathrm{i}\omega T}\right)$, and by $\mathbf{a}_i\left(e^{\mathrm{i}\omega T}\right)$ the length-$R$ column vector of $\left\{A_{i,r}\left(e^{\mathrm{i}\omega T}\right)\right\}_{r=1,\cdots,R}$ for the $i$th data stream, we can write
\begin{align}\label{Fourier_Samples_UoS}
	\bdsb{z}\left(e^{\mathrm{i}\omega T}\right) &=
	\mathbf{M}_{\phi\phi}(\omega,\bdsb{k},\bdsb{q})\mathbf{a}\left(e^{\mathrm{i}\omega T}\right)+\mathbf{v}\left(e^{\mathrm{i}\omega T}\right)
\end{align}
over the domain $\omega\in[-\pi/T,\pi/T]$. The derivation is identical to the development in \citep{Eldar_time_delay} and is therefore omitted. Here $\mathbf{a}\left(e^{\mathrm{i}\omega T}\right)\triangleq[\mathbf{a}_1^H\left(e^{\mathrm{i}\omega T}\right),\cdots,\mathbf{a}_I^H\left(e^{\mathrm{i}\omega T}\right)]^H$ is a length-$IR$ vector containing the DTFT of all the data sequences $\{a_{i,r}[n]\}_{n\in\mathbb{Z}}$ and $\mathbf{M}_{\psi\phi}(\omega,\bdsb{k},\bdsb{q})$ is an $I|\mathcal{K}||\mathcal{Q}|\times IR$ matrix with $\left[(i,k,q),(i,r)\right]$th element
\begin{align}\label{M_GPS}
	\left[\mathbf{M}_{\phi\phi}(\omega,\bdsb{k},\bdsb{q})\right]_{(i,k,q),(i,r)} &=
	\frac{1}{T}\sum_{\ell=0}^{LM-1} \Phi_{i,k,q}^\ast\left(\omega-\frac{2\pi \ell}{T}\right)\Phi_{i,k_{i,r},q_{i,r}}\left(\omega-\frac{2\pi \ell}{T}\right).
\end{align}
The component $\mathbf{v}\left(e^{\mathrm{i}\omega T}\right) = [\cdots,v_{i,k,q}\left(e^{\mathrm{i}\omega T}\right),\cdots]^T$ is the filtered noise by matched filters (generators) $\{\phi_{i,k,q}(t)\}_{i=1,\cdots,I}^{k\in\mathcal{K},q\in\mathcal{Q}}$ and therefore has a cross-spectral density matrix $\mathbf{R}_{vv}\left(e^{\mathrm{i}\omega T}\right) = \sigma^2\mathbf{M}_{\phi\phi}(\omega,\mathcal{K},\mathcal{Q})$, where $\mathbf{M}_{\phi\phi}(\omega,\mathcal{K},\mathcal{Q})$ is the Gram matrix of all the generators defined by
\begin{align}
	\left[\mathbf{M}_{\phi\phi}(\omega,\mathcal{K},\mathcal{Q})\right]_{(i',k',q'),(i,k,q)}  &= \frac{1}{T}\sum_{\ell=0}^{LM-1}\Phi_{i',k',q'}^\ast\left(\omega-\frac{2\pi \ell}{T}\right)\Phi_{i,k,q}\left(\omega-\frac{2\pi \ell}{T}\right).
\end{align}

Exploiting the specific choice of sampling kernels and the structure of $\mathbf{M}_{\phi\phi}(\omega,\bdsb{k},\bdsb{q})$ and $\mathbf{M}_{\phi\phi}(\omega,\mathcal{K},\mathcal{Q})$, we can further analyze the output samples $\bdsb{z}\left(e^{\mathrm{i}\omega T}\right)$ as stated below.
\begin{thm}\label{theorem1}
	Suppose that the following conditions hold,
	\begin{itemize}
	\item[{\bf C1)}] the pulse shaping filter has a spectrum $G(\omega) = \left[1+ \mathcal{\epsilon}(\omega)\right]\mathrm{rect}_{2\pi L/T_c}(\omega)$ with error $\epsilon_g(\omega)$;
	\item[{\bf C2)}] the C/A code cross spectral density is $S_{i'i}\left(e^{\mathrm{i}\omega T_c}\right)= \delta[i'-i]+\epsilon_{i',i}(\omega)$ with error $\epsilon_{i',i}(\omega)$;
	\item[{\bf C3)}] the frequency search step size is chosen as $\Delta\omega = 2\pi j/T$ and $j\in\mathbb{Z}^{+}$.
	\end{itemize}
	If the error functions satisfy $\|\epsilon_g(\omega)\|\ll 1$ and $\|\epsilon_{i',i}(\omega)\|\ll 1$ for any $i',i=1,\cdots,I$, then the Gram matrix of all the generators $\{\phi_i(t-qT_c)e^{\mathrm{i}k\Delta\omega t}\}_{i=1,\cdots,I}^{k\in\mathcal{K},q\in\mathcal{Q}}$ satisfies
	\begin{align}
		\mathbf{M}_{\phi\phi}(\omega,\mathcal{K},\mathcal{Q}) =LM \mathbf{I} +\mathbf{E}(\omega),
	\end{align}
	where $\mathbf{E}(\omega)$ is bounded perturbation matrix satisfying $\|[\mathbf{E}(\omega)]_{(i',k',q'),(i,k,q)}\|=\mathcal{O}(1)\ll LM$ and the filtered noise samples have a cross-spectral density matrix $\mathbf{R}_{vv}\left(e^{\mathrm{i}\omega T}\right)=\sigma^2 [LM\mathbf{I}+\mathbf{E}(\omega)]$. The output samples $\bdsb{z}[n]=\left[\cdots,z_{i,k,q}[n],\cdots\right]^T$ at each of the kernels $\phi_{i,k,q}(t)$ can be written as
	\begin{align*}
		z_{i,k,q}[n] =
		\begin{cases}
			LMa_{i,r}[n] +\mathcal{O}(1) + v_{i,k,q}[n], & q=q_{i,r}~\mathrm{and}~k=k_{i,r}\\
			\mathcal{O}(1) + v_{i,k,q}[n],  & \mathrm{otherwise},
		\end{cases}
	\end{align*}
	where $v_{i,k,q}[n]$ is the time-domain filtered noise sample and $\mathcal{O}(1)\ll LM$ is some bounded perturbation error with $LM$ being the processing gain on the signal-to-noise ratio.
\end{thm}
\begin{proof}
See Appendix A.
\end{proof}
Note that the frequency step size $\Delta\omega=2\pi j/T$ corroborates the fact that for standard commercial GPS systems, the step size is usually $2\pi \times 500$ rads/s which fits the analysis here by choosing $ j =10$. Also, we can see that the output $z_{i,k,q}[n]$ at each sampler represents the correlation between the matched filters and the incoming signal, which is proportional to the magnitude of $a_{i,r}[n]$ and corrupted by noise. Assuming large enough processing gain $LM$ and small enough noise, the delay-Doppler pairs $\{\tau_{i,r}=q_{i,r}T_c\}_{i=1,\cdots,I}^{r=1,\cdots,R}$ and $\{\omega_{i,r}=k_{i,r}\Delta\omega\}_{i=1,\cdots,I}^{r=1,\cdots,R}$ can be found by the location of the peaks/dominant entries in $z_{i,k,q}[n]$. The strongest set of satellite signals can then be detected by comparing the values in $z_{i,k,q}[n]$ so that a subset $\mathcal{I}$ of the satellite signals are locked and passed onto the tracking stage for finer extraction. If we ignore the noise for a moment, then $z_{i,k,q}[n]$ is sparse in the sense that for each value $n$ it contains only a small number of non-zero entries.

\subsection{Compressive Multichannel Acquisition}
Although effective, this conventional approach taken by standard GPS receivers performs exhaustive correlations (MF approach) that requires abundant samples from a large number of correlators $I|\mathcal{Q}||\mathcal{K}|$. This task can be computationally expensive and  demanding on the hardware and memory resources. Assuming a maximum channel delay spread of $\tau_{\rm max}=QT_c$ and Doppler shift of $|\omega_{\rm max}| = K\Delta\omega$, the total number of correlators is $2IQK$. For example, the maximum Doppler shift is typically $\pm 10$kHz. Assuming a delay spread up to code length $\tau_{\rm max}=MT_c$, then with a frequency grid of $500$Hz, the total number of correlators needed becomes $24 \times 1023 \times 42\approx  10^{6}$.

Therefore, it is highly desirable to scale down the computational complexity and power consumption of a user GPS device by performing less correlations while sustaining its capability to pinpoint the signal timing and Doppler information during acquisition. By observing the correlation outputs in the vector $\bdsb{z}[n]=[\cdots,z_{i,k,q}[n],\cdots]^T$, it can be seen that only few of the dominant entries are useful. Our goal is to exploit the underlying sparsity in the signal model to design an acquisition scheme that requires far fewer correlators. Instead of tackling the problem from a match-filtering viewpoint as in standard GPS, we look at the problem from an analog CS perspective \citep{Eldar_CS}, which is one of the main contributions of this paper.

The analog CS design outlined in \citep{Eldar_CS} requires a small number of samplers (only twice the sparsity $2|\mathcal{I}|R$ in a noiseless setting), and hence gives rise to substantial practical savings as analyzed later in Section 7. However, the solution \citep{Eldar_CS} is given in the frequency domain and in general does not admit a tractable form in time domain, which makes it hard to implement in practice. Another contribution of this work lies in further exploiting the structure of GPS signals so that the sampling kernels are easy to implement. The outputs from the compressive samplers can then be used to solve the sparse recovery problem of locating the dominant/peak values reflected in the vector $\bdsb{z}[n]$, for example, using the method in \citep{Eldar_MMV}.

Before we go into the details of our design, we start by describing the analog CS framework \citep{Eldar_CS}. In Section 5 we further develop and simplify the general solution to fit our problem.

\section{Compressed Sensing of Analog Signals}
The exhaustive MF approach in standard GPS receivers acquires delays and Dopplers by directly exposing the sparse structure in the output samples $\bdsb{z}[n]$ obtained from a large number of correlators. In order to reduce the number of correlators while retaining the ability to correctly identify the peaks of $\bdsb{z}[n]$, it is possible to directly measure a compressed version of $\bdsb{z}[n]$ at the samplers outputs and recover that sparse structure instead, by employing analog CS techniques.
\subsection{General Model for Analog Compressed Sensing (CS)}
The signal model in (\ref{discretized_model}) does not reflect any sparse structure, since it is expressed by a set of deterministic generators $\phi_i(t)$'s defined by unknown parameters $q_{i,r}$ and $k_{i,r}$. The sparsity we exploit is the sparsity of delay-Doppler pairs pinpointed by the peaks/dominant entries in $\bdsb{z}[n]$ over the entire delay-Doppler space $\mathcal{Q}\times\mathcal{K}$ for each user $i=1,\cdots,I$ that is informative in acquiring the signal. Using a dictionary $\{\phi_i(t-qT_c)e^{\mathrm{i}k\Delta\omega t}\}_{k\in\mathcal{K},q\in\mathcal{Q}}$, the signal can be equivalently expressed by
\begin{align*}
	x(t)	= \sum_{n\in\mathbb{Z}} \sum_{i=1}^{I}
	     \sum_{k\in\mathcal{K}}\sum_{q\in\mathcal{Q}}y_{i,k,q}[n]\phi_i(t-nT-qT_c)e^{\mathrm{i}k\Delta\omega t} +v(t),\nonumber
\end{align*}
where
\begin{align}\label{sparse_model}
	y_{i,k,q}[n] =
	\begin{cases}
		a_{i,r}[n], & q=q_{i,r}~\mathrm{and}~k=k_{i,r}\\
		0, & \mathrm{otherwise}.
	\end{cases}
\end{align}
Note that the sparsity of $\bdsb{y}[n]\triangleq [\cdots, y_{i,k,q}[n],\cdots]^T$ is identical to that of $\bdsb{z}[n]$ in the noiseless setting. Indeed, for each $i=1,\cdots,I$ there are altogether $R$ dominant coefficients $\{y_{i,k,q}[n]\}_{k\in\mathcal{K},q\in\mathcal{Q}}$ that correspond to the original coefficients $\{a_{i,r}[n]\}_{r=1,\cdots,R}$ and select the correct code-phase $q_{i,r}$ and Doppler shifts $k_{i,r}$. Let the support of $\bdsb{y}[n]$ be $\mathcal{S}$, then the support $\mathcal{S}$ contains the code-phase and Doppler information for acquisition, with a sparsity of $|\mathcal{S}|=|\mathcal{I}|R$. The aim of analog CS is to exploit this sparsity in acquiring $x(t)$ using fewer correlators.

\begin{figure}
\center
\includegraphics[width=4.5in]{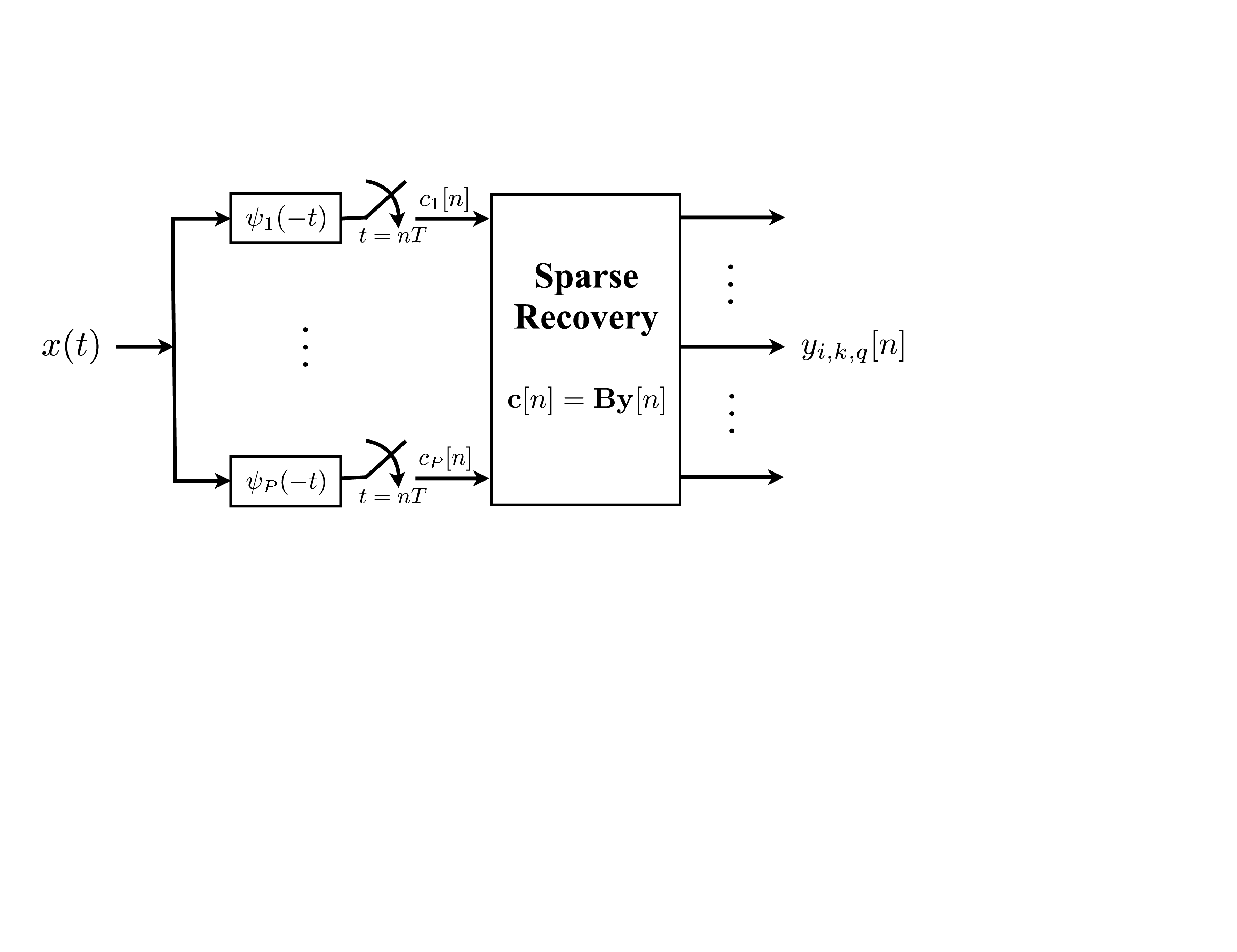}
\caption{Compressive Multichannel Sampling (CS)}
\end{figure}

\subsection{General Solution of Compressive Samplers}
As the scheme of \citep{Eldar_CS} uses a set of compressive samplers $\psi_p(-t)$, $p=1,\cdots, P\ll I|\mathcal{K}||\mathcal{Q}|$ to obtain minimal measurements, from which the sparse vector $\bdsb{y}[n]$ can be recovered. As depicted in Fig. 2, the samples at the output of $\psi_p(-t)$ at $t=nT$ are given by 
\begin{align}
	c_p[n]\triangleq \langle x(t), \psi_p(t-nT) \rangle.
\end{align}
Similar to the mathematical manipulations in Section 3, the system equation can be re-written as
\begin{align}\label{CS_model}
	\mathbf{c}\left(e^{\mathrm{i}\omega T}\right) &=
	 \mathbf{M}_{\psi\phi}(\omega,\mathcal{K},\mathcal{Q})\bdsb{y}\left(e^{\mathrm{i}\omega T}\right)+\mathbf{w}\left(e^{\mathrm{i}\omega T}\right),
\end{align}
where $\mathbf{M}_{\psi\phi}(\omega,\mathcal{K},\mathcal{Q})$ is a $P\times I|\mathcal{K}||\mathcal{Q}|$ matrix with similar structure to (\ref{M_GPS}) and the notation $\mathbf{w}\left(e^{\mathrm{i}\omega T}\right)$ is used to distinguish the noise component from the previous method in standard GPS. It has been proven in \citep{Eldar_CS} that in a noiseless setting, simply twice the sparsity $P=2|\mathcal{I}|R$ is needed for successful recovery of the sparse vector $\bdsb{y}[n]$, if $\psi_p(-t)$'s are chosen properly. For noisy scenarios, the necessary number of channels $P$ is larger than the minimum, and evaluated numerically; in any case, it is much smaller than that required by the standard scheme, as we will demonstrate in Section 6.

This reduction is obtained by appropriately choosing a set of randomized correlators $\bdsb{\Psi}(\omega)\triangleq [\Psi_1(\omega),\cdots,\Psi_P(\omega)]^T$. A general expression of the compressive samplers is given in \citep{Eldar_CS} as
\begin{align}\label{CS_kernel}
	\bdsb{\Psi}(\omega) = \mathbf{B}\mathbf{M}_{\phi\phi}^{-1}(\omega,\mathcal{K},\mathcal{Q})\bdsb{\Phi}(\omega,\mathcal{K},\mathcal{Q}),
\end{align}
where $\mathbf{B}$ is a sensing matrix satisfying certain coherence properties \citep{Tao} (e.g., Gaussian random matrix or partial DFT matrix \citep{Tao}, or an appropriate deterministic binary matrix \citep{Calderbank}), and $\bdsb{\Phi}(\omega,\mathcal{K},\mathcal{Q})\triangleq [\cdots, \Phi_i(\omega-k\Delta\omega)e^{-\mathrm{i}(\omega-k\Delta\omega)qT_c},\cdots]^T$ is a length-$I|\mathcal{K}||\mathcal{Q}|$ vector containing the Fourier transforms of the generators $\{\phi_i(t-qT_c)e^{\mathrm{i}k\Delta\omega t}\}_{i=1,\cdots,I}^{k\in\mathcal{K},q\in\mathcal{Q}}$. With this choice of $\bdsb{\Psi}(\omega)$, it can be shown that $\mathbf{M}_{\psi\phi}(\omega,\mathcal{K},\mathcal{Q})=\mathbf{B}$. Since $\mathbf{B}$ is independent of frequency $\omega$,  transforming \eqref{CS_model} into the time domain, the samples can be written as
\begin{align}\label{IMV}
	\mathbf{c}[n] = \mathbf{B}\bdsb{y}[n]+\mathbf{w}[n],\quad n\in\mathbb{Z}.
\end{align}
The vectors $\{\bdsb{y}[n]\}$ are jointly sparse since they all share the same sparsity pattern. To find $\bdsb{y}[n]$, we can convert \eqref{IMV} to a finite MMV problem using the continuous-to-finite (CTF) technique developed in \citep{Eldar_MMV}. Specifically, we first find a basis for the range space of $\{\mathbf{c}[n]\}$ by computing the covariance matrix $\mathbf{R}_{\mathbf{c}\mathbf{c}}$ and decomposing it as $\mathbf{R}_{\mathbf{c}\mathbf{c}} = \mathbf{C}\mathbf{C}^H$. Here $ \mathbf{C}$ can be chosen as the eigenvectors of $\mathbf{R}_{\mathbf{c}\mathbf{c}}$ multiplied by the square-root of the corresponding eigenvalues. Then, the support of $\bdsb{y}[n]$, $n\in\mathbb{Z}$ can be obtained by  solving $\mathbf{C}=\mathbf{B}\mathbf{Y}$, where $\mathbf{Y}$ is the sparsest matrix satisfying the measurement equation. This problem can be treated using various MMV sparse recovery techniques \citep{Cotter}\citep{Chen}. In our simulations, we use the ReMBo algorithm developed in \citep{Eldar_MMV}. Finally the support of $\bdsb{y}[n]$ is obtained by taking the union of the supports of the columns in the matrix $\mathbf{Y}$. Once the support of $\bdsb{y}[n]$ is recovered, the acquisition of correct delay-Doppler pair is automatically achieved by locating the dominants/peaks in the vector $\bdsb{y}[n]$ of (\ref{sparse_model}).

{\bf Remark:} As verified in Section 6, as the number of correlators $P$ increases, the acquisition performance improves significantly. Solving the MMV problem requires collecting multiple measurement vectors $\{\mathbf{c}[n]\}$, while  the standard GPS scheme can either employ information for a single measurement $\bdsb{z}[n]$ in (\ref{z}) or further leverage the processing gain over multiple measurements $\{\bdsb{z}[n]\}$. For the proposed compressive acquisition scheme, if a single vector measurement is used to recover the sparse vector $\bdsb{y}[n]$ using greedy methods or $\ell_1$-norm based methods, the performance will degrade as shown in Fig. \ref{fig1:sm_offtg_uid} but not significantly. Therefore, there is a trade-off between the number of observations $\mathbf{c}[n]$, the number of acquisition channels $P$ as well as the accuracy of the acquisition in comparison with the standard GPS scheme.

\section{Simplified Randomized Correlators}
The method proposed in \citep{Eldar_CS} depends on the ability of physically implementing the sampling kernels in (\ref{CS_kernel}). Therefore, we explore the structure of the matrix $\mathbf{M}_{\phi\phi}(\omega,\mathcal{K},\mathcal{Q})$ to provide practical insights on the design of such filters.
\begin{corr}
Suppose that the conditions $(\mathbf{C1})$-$(\mathbf{C3})$ and the requirement on the error functions in Theorem 1 hold. Then the sampling kernels can then be chosen as the randomized correlators
\begin{align}\label{filter_response}
	\psi_p(t) = \sum_{i=1}^{I}\sum_{k\in\mathcal{K}}\sum_{q\in\mathcal{Q}}b_{p,(i,k,q)}\phi_i(t-qT_c)e^{\mathrm{i} k\Delta\omega t}, \quad p=1,\cdots, P.
\end{align} 
\end{corr}
\begin{proof}
From \eqref{CS_kernel} we have the general solution of the compressive samplers
\begin{align}
	\bdsb{\Psi}(\omega) = \mathbf{B}\mathbf{M}_{\phi\phi}^{-1}(\omega,\mathcal{K},\mathcal{Q})\bdsb{\Phi}(\omega,\mathcal{K},\mathcal{Q}).
\end{align}
According to the result in Theorem 1, using Taylor expansion on the matrix inverse $\mathbf{M}_{\phi\phi}^{-1}(\omega,\mathcal{K},\mathcal{Q})$ and ignoring high order terms scaled by $1/LM \ll 1$, we can approximate the inverse by
\begin{align}
	\left(\mathbf{I}+\frac{1}{LM}\mathbf{E}(\omega)\right)^{-1}
	&=
	\mathbf{I} - \frac{1}{LM}\mathbf{E}(\omega) + \frac{1}{(LM)^2}\mathbf{E}^2(\omega) - \frac{1}{(LM)^3}\mathbf{E}^3(\omega) \cdots
	\approx \mathbf{I},
\end{align}
where the last approximation comes from the fact that $\mathbf{E}(\omega)$ contains negligible elements. Therefore, the compressive samplers can be chosen directly as $\bdsb{\Psi}(\omega) = \mathbf{B}\bdsb{\Phi}(\omega,\mathcal{K},\mathcal{Q})$, which leads to the time-domain expression in the corrolary.
\end{proof}
The The filter responses of \eqref{filter_response} can be precomputed, and these $P$ channel outputs are sampled every $T=MT_c$ to produce the test statistics that are going to be used in lieu of the coefficients $\bdsb{z}[n]$ in Theorem 1.

%%%%%%%%%%%%%%
\begin{algorithm}[t]
{\small
\caption{Compressive Multichannel Acquisition} \vspace{1ex}
(1) Construct $P$ compressive sampling kernels as
\begin{align}
	 \psi_p(t)=\sum_{i=1}^{I}\sum_{k\in\mathcal{K}}\sum_{q\in\mathcal{Q}}b_{p,(i,k,q)}\phi_i(t-qT_c)e^{\mathrm{i} k\Delta\omega t},
\end{align}
where $[\mathbf{B}]_{p,i,k,q}=b_{p,i,k,q}$ is a sensing matrix that satisfies certain coherence properties such as RIP \citep{Tao}.\\
(2) Apply the set of compressive sampling kernels $\bdsb{\Psi}(\omega)$ and arrive at measurements
\begin{align*}
	\mathbf{c}[n] &= \mathbf{B}\bdsb{y}[n] + \mathbf{w}[n], \quad n\in\mathbb{Z}.
\end{align*}
(3) Solve the jointly sparse recovery problem as in \citep{Eldar_MMV} to recover the support of $\bdsb{y}[n]$.

(4) Once the support of $\bdsb{y}[n]$ is available, the delay-Doppler pairs are determined by the support $q=q_{i,r}$ and $k=k_{i,r}$ as in (\ref{sparse_model}).\\
}
\end{algorithm}
%%%%%%%%%%%%%%

{\bf Remark}: Note that although the samples are taken at $1/T$, the physical implementation of the compressive multichannel filtering operation is likely to require digital processing at the chip rate $1/T_c$. Nevertheless, it is possible that a wise choice of the coefficients of the matrix $\mathbf{B}$ can further help reduce computations while maintaining the identifiability of the parameters. Analysis of this approach goes beyond our current scope.  What we can certainly claim is that the number of computations is now controlled by the parameter $P$, rather than by the number of possible generators that span all possible delays $\mathcal{Q}$ and Dopplers $\mathcal{K}$. In fact, the sampling Kernels are precomputed and used online. This is likely to reduce cost of computation, access to memory and storage. The performance of the compressive multichannel sensing structure degrades gracefully as $P$ decreases, giving designers degrees of freedom to choose a desirable operating point.

%%%%%%%%%%% Numerical Results external file

\section{Numerical Results}
In this section, we run numerical simulations to demonstrate the proposed CS acquisition scheme in GPS receivers. In the simulation, $|\mathcal{I}|=4$ out of $I=24$ satellites asynchronously transmit C/A signals that are received by the GPS devices, where the codes are length-$M=NM_0$ Gold sequences with $N=20$ and $M_0=1023$. A total of $n=50$ navigation data bits are sent at the rate of $1/T=50$Hz (i.e., $T=20$ms).The transmit filter is modeled by a finite length pulse shaping filter $g(t)=\sqrt{T_c}\mathrm{sinc}(t/T_c)$ when $|t| \leq T_g$, and $g(t)=0$ otherwise. The length $T_g$ is sufficiently large so that the response of the pulse in the frequency domain remains approximately flat, i.e. $G (\omega) \approx \mathrm{rect}_{2\pi/T_c}(\omega)$.

To reduce the simulation overhead without incurring a loss of generality, we assume that our statistical model for the channel consists of uniformly distributed delays, $\tau_{i,r} \thicksim \mathcal{U}(0,\tau_{\rm max})$ that are bounded by a maximum delay spread of $\tau_{\rm max}=20T_c$; and of Doppler shifts, $|\omega_{i,r}| \leq \omega_{\rm max}$ that are uniformly distributed, $\omega_{i,r} \thicksim \mathcal{U}(-\omega_{\rm max},\omega_{\rm max})$ over a frequency range delimited by $\omega_{\rm max}/2\pi= 2.5$kHz. The channel gains are $h_{i,r} \thicksim \mathcal{CN}(0,1)$, with a multi-path propagation having $R=2$ paths per satellite. In order to identify fractional delays with a half-chip accuracy, the functions $\phi_{i,k,q}(t)=\phi_i(t-qT_c)e^{\mathrm{i}k\Delta\omega t}$ are chosen with a half-chip spacing $q=0,1/2,1,\cdots$ such that the resolution of $\Delta\tau=T_c/2$ is achieved, and with a frequency resolution of $\Delta \omega = 10 \times 2\pi/T$ that corresponds to steps around $500$Hz when $T=20$ms. It follows that $|\mathcal{Q}|=\lceil \tau_{\rm max}/\Delta \tau\rceil+1=41$ and $|\mathcal{K}|=2 \lceil\omega_{\rm max}/\Delta \omega\rceil+1=11$. For simulation purpose, the sensing matrix $\mathbf{B}$ is generated as a random binary matrix (while in practice it can be chosen as a deterministic binary matrix to simplify the implementation of correlators \citep{Calderbank}).

%
%% satellite identification
\begin{figure}[t]
\begin{minipage}[b]{0.5\linewidth}
\centering
\includegraphics[width=2.5in,height=3in]{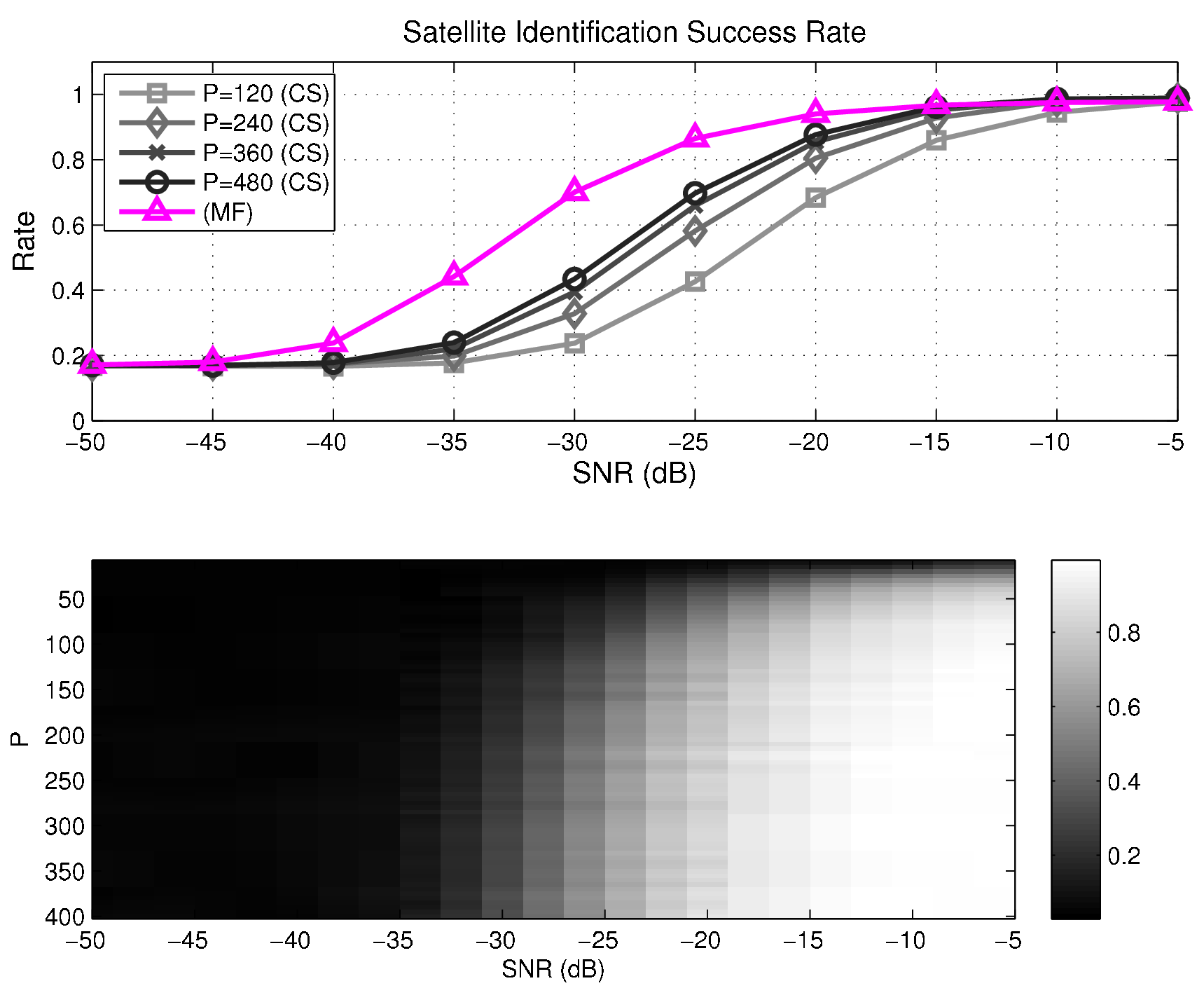}
\caption{\small Satellite identification rate $\mathbb{P}(\mathcal{\widehat{I}} = \mathcal{I})$ using a single measurement $\mathbf{c}[1]$  for a CS receiver v.s. the MF receiver for $P=\{120,240,360,480\}$ (above), and for $P=\{10,20,\ldots,400\}$ (below)}

\label{fig1:sm_offtg_uid}
\end{minipage}
\hspace{0.5cm}
\begin{minipage}[b]{0.5\linewidth}
\centering
\includegraphics[width=2.5in,height=3in]{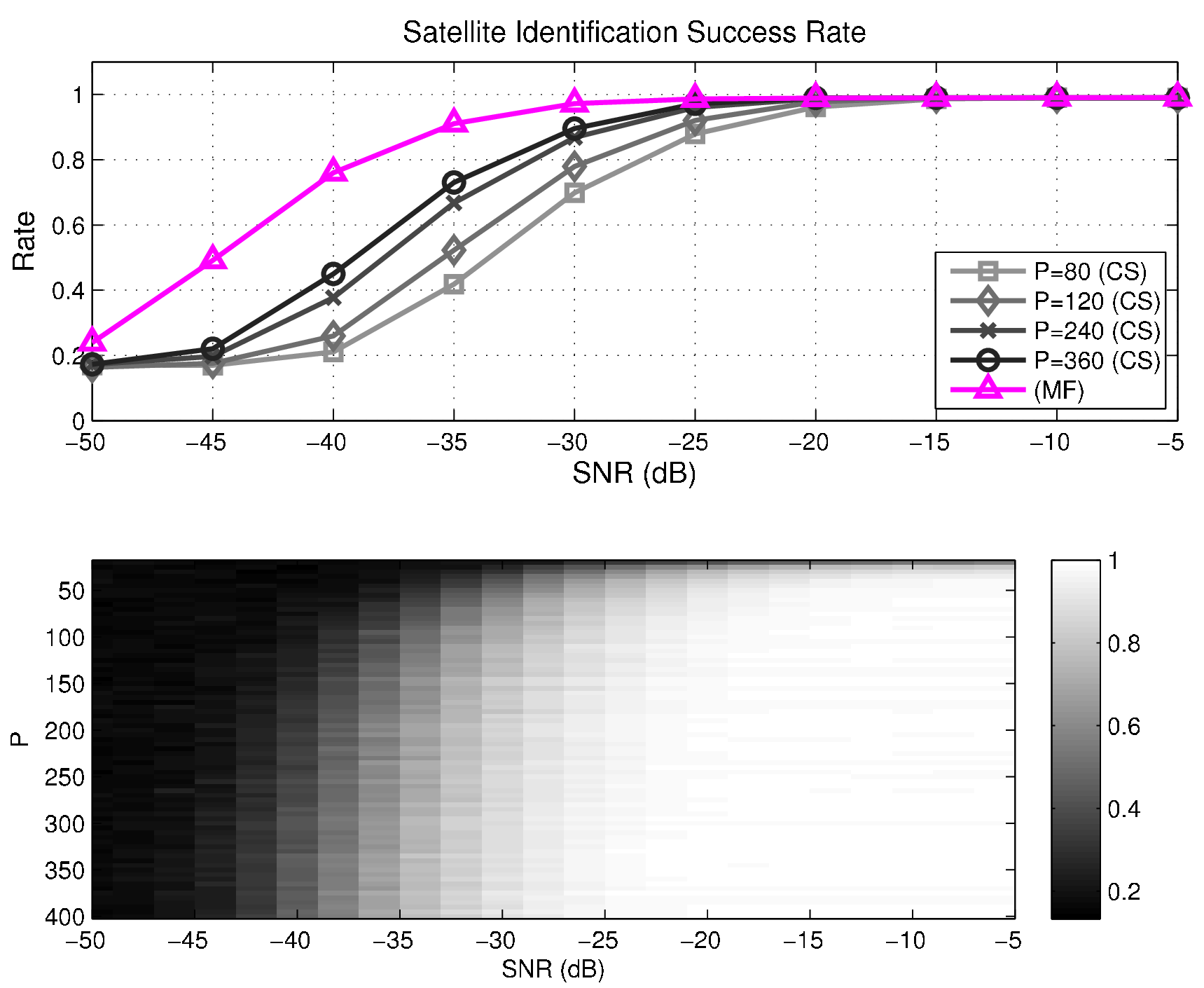}
\caption{\small Satellite identification rate $\mathbb{P}(\mathcal{\widehat{I}} = \mathcal{I})$ using multiple measurements $\{\mathbf{c}[1],\mathbf{c}[2],\ldots,\mathbf{c}[50]\}$ for CS receiver v.s. the MF receiver for $P=\{80,120,240,360\}$ (above), and for $P=\{10,20,\ldots,400\}$ (below)}
\label{fig2:mm_offtg_uid}
\end{minipage}
\end{figure}
%% rmse identification
\begin{figure}[t]
\begin{minipage}[b]{0.5\linewidth}
\centering
\includegraphics[width=2.5in,height=3in]{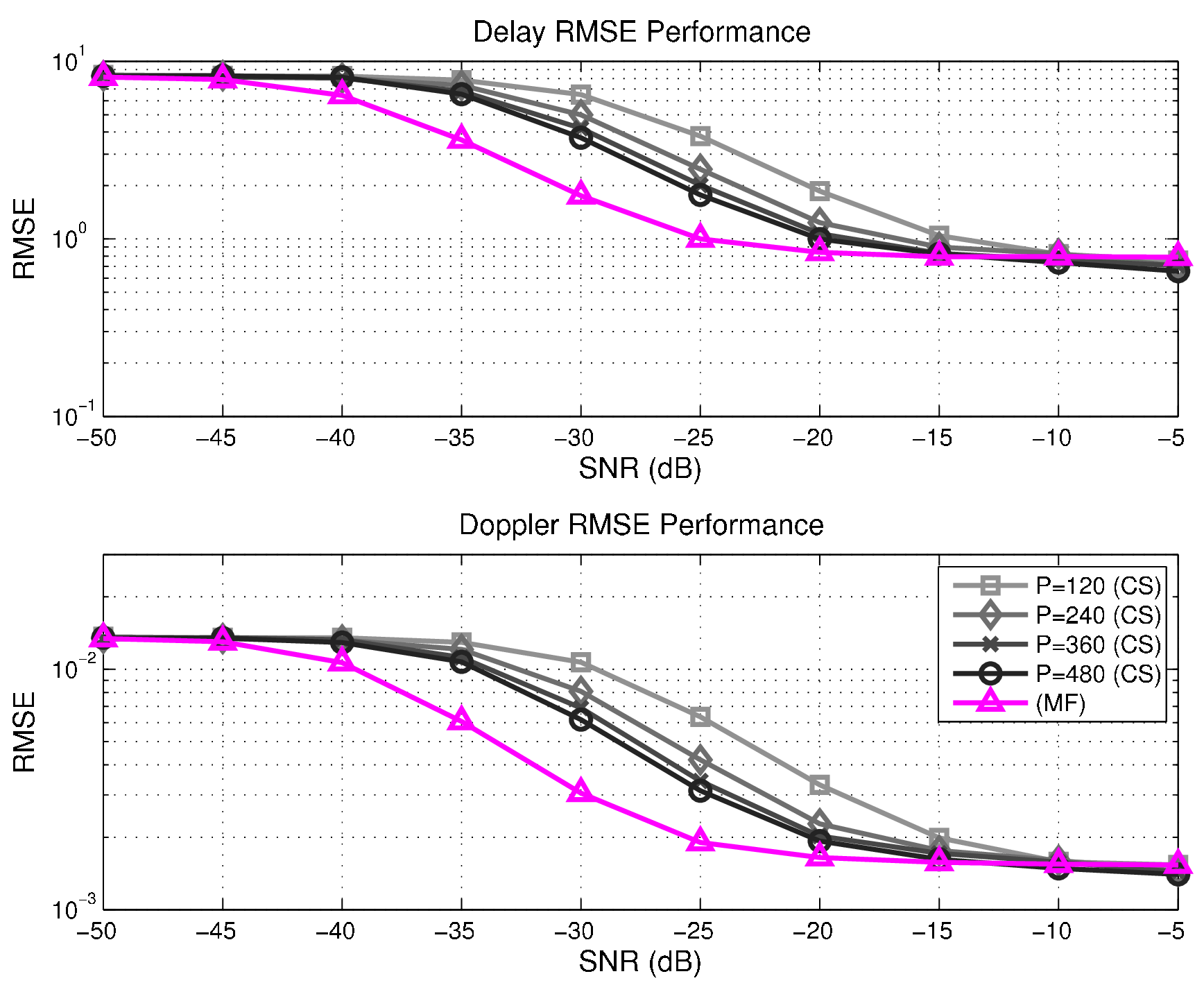}
\caption{\small Delay estimation (above) and Doppler estimation (below) performance of the CS, with $n=1$ and $P=\{120,240,360,480\}$ compared against the the MF receiver}
\label{fig3:sm_offtg_mse}
\end{minipage}
\hspace{0.5cm}
\begin{minipage}[b]{0.5\linewidth}
\centering
\includegraphics[width=2.5in,height=3in]{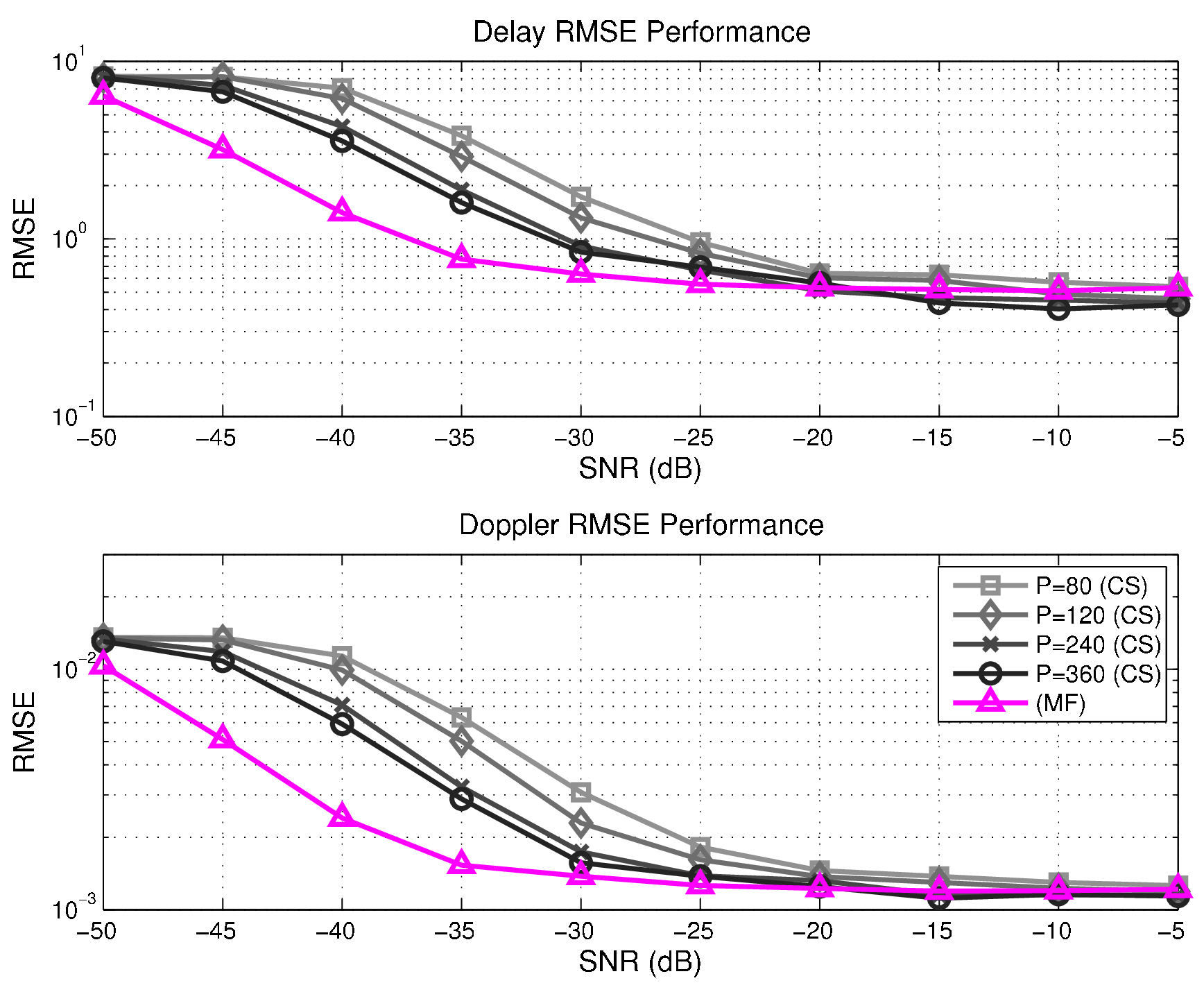}
\caption{\small Delay estimation (above) and Doppler (estimation) performance of the CMA, with multiple measures ($n=50$) and $P=\{80,120,240,360\}$, compared against the the MF receiver also processing $n=50$ measures}
\label{fig4:mm_offtg_mse}
\end{minipage}
\end{figure}

In all simulations, the attenuated components with distinct delays from each of the satellites are acquired by a number of $P=\{80, 120, 240, 360, 480\}$ channels, in contrast to the traditional $24\times 41 \times 11 \approx 1 \times 10^4$. The performance is illustrated in terms of success rate and average Root Mean Square Error (RMSE), respectively, in Fig. \ref{fig1:sm_offtg_uid} and Fig. \ref{fig2:mm_offtg_uid}. The success rate of acquisition is the probability $\mathbb{P}(\widehat{\mathcal{I}} = \mathcal{I})$ of the proposed scheme to determine the strongest $|\mathcal{I}|=4$ signals, which is shown in the figure against the number of channels $P$ and the SNR. The conditional RMSE is an average error between the true delay-frequency parameters and those associated to the strongest paths of the correctly identified satellites

\begin{equation}\label{numres:eq1}
\mathrm{RMSE}_{\rm average} (\bdsb{q}) \triangleq \sqrt{\frac{1}{|\{ \widehat{\mathcal{I}} \cap \mathcal{I} \} |}\sum_{i\in \{ \widehat{\mathcal{I}} \cap \mathcal{I} \} } (\widehat{q}_{i}\Delta \tau-\tau_{i})^2}, \nonumber
\end{equation}
where $(\tau_i,\omega_i) \triangleq ( \tau_{i,r^{\ast}},\omega_{i,r^{\ast}})$ with $r^{\ast}=\argmax_{r \in \{1,\ldots,R\}}| h_{i,r}  |^2$ and
\begin{equation}
( \widehat{q}_i ,\widehat{k}_i ) = \argmax_{q \in \mathcal{Q},k \in \mathcal{K}} |z_{i,k,q}[n]|^2 = \argmax_{q \in \mathcal{Q},k \in \mathcal{K}} |y_{i,k,q}[n]|^2
\end{equation}
are the delay-frequency index pairs of strongest path associated to the $i$th satellite. Similarly, the average RMSE for the Doppler is
\begin{equation}
\mathrm{RMSE}_{\rm average} (\bdsb{k}) \triangleq \sqrt{ \frac{1}{|\{\widehat{\mathcal{I}} \cap \mathcal{I} \} |}\sum_{i \in \{ \widehat{\mathcal{I}} \cap \mathcal{I} \} } (\widehat{k}_{i}\Delta \omega-\omega_{i})^2 }, \nonumber
\end{equation}
Although the compressive acquisition scheme suffers from a compression loss, both Fig.\ref{fig1:sm_offtg_uid} and Fig.\ref{fig2:mm_offtg_uid} highlight its ability to perform closely as the traditional MF. When $P \geq 80$ and $\textrm{SNR} \geq -25$ dB the active satellites $\mathcal{I}$ can be identified satisfactorily which leads to great savings (less than $1$\% of the original $1\times 10^4$).

The figures above illustrate acquisition performances using a single set of measurements $\mathbf{c}[0]$ against that using multiple sets of measurements $\{\mathbf{c}[n]\}_{n=1}^{50}$. Using a single measurement suffers from a performance loss ($-10$ dB for $P=120$ at the rate of approximately $0.8$). In fact, by reducing $n$, the accuracy of $\bdsb{z}[n]$ and consequently the sensitivity, degrade. Furthermore, it can be seen from Fig. \ref{fig1:sm_offtg_uid} that the required number of channels $P$ has to be raised to $480$ (less than $5$\% of the original $1\times 10^4$) to achieve a reliable rate that approaches the MF result.

A similar trend is also visible on the conditional RMSE curves for both single (Fig. \ref{fig3:sm_offtg_mse}) and multiple (Fig. \ref{fig4:mm_offtg_mse}) modes ($-12$ dB for $P=120$ when $\text{RMSE}(\bdsb{q}) \approx 2$ and $\text{RMSE}(\bdsb{k}) \approx 2 \cdot 10^{-3}$). At high SNR the performance is limited by the presence of a systematic error due to the modeling mismatch from the quantized parameters. At low SNR, instead, the error is bounded by the length of the search interval $QT_c$. Once again the CS method closely approaches the MF performance, especially when $n=1$.

\section{Complexity Analysis}
The complexity of the acquisition algorithm is due to two aspects: 1) {\it storage requirement} and 2) {\it computational complexity}. We provide here a brief analysis of the complexity of the proposed CS scheme against traditional MF scheme. To make a fair and practical comparison, we assume that the implementation is done in the digital domain and we use the $L_{\rm kernel}$-tap digitized version of the sampling kernels $\{\psi_p(t)\}_{p=1}^P$ (and also $\{\phi_{i,k,q}(t)\}_{i=1,\cdots,I}^{k\in\mathcal{K},q\in\mathcal{Q}}$ for the traditional case).\\
\subsection{Storage and Processing Requirement}
The difference in storage results from two sources, one is the storage for the digital kernel taps and the other is the outputs of the sampling kernels used for peak recovery, both of which are proportional to the number of sampling kernels. Furthermore, the processing overhead per unit of time for these stored values scales proportionally with the storage requirement as well.
	\begin{center}
		\begin{tabular}{ l*{3}{l}r}
        \toprule
        \toprule
    	&  sampling kernels     				& output samples \\
		\hline \textbf{CS Receiver}  &  $P\times L_{\rm kernel}$		       			 &  $\mathcal{O}(P)$ \\
		\hline \textbf{MF Receiver}    &  $I|\mathcal{K}||\mathcal{Q}|\times L_{\rm kernel}$   &  $\mathcal{O}(I|\mathcal{K}||\mathcal{Q}|)$ \\
		\hline 	
		\end{tabular}
	\end{center}
    It is clear that the proposed compressive acquisition scheme handles less data, which facilitates the pipelining of the algorithm and also relieves the burden of storage.\\
\subsection{Computational Complexity} 
The difference in computations stems from the correlations and the search for the peak. The number of operations in performing correlations is proportional to the number of sampling kernels, while the peak recovery is different for the two approaches, depending on how the sparse recovery (proposed CS structure) and the exhaustive search (MF structure) are implemented. Here we further compare the two architectures by their number of operations that are necessary to identify the delay-Doppler pairs \eqref{numres:eq1}. In this practical analysis, the compressed samples $\mathbf{c}[n]$ are obtained by post-processing of the digitally sampled versions of $x(t)$ at the chip rate and processed using a greedy algorithm Orthogonal Matching Pursuit (OMP) \citep{Omp93}. Note that using analog implementation in the acquisition can further bring down the complexity in terms of processing.

We introduce a vector $\mathbf{x}[n]$ of $M$ dimensions, whose $m$th entry is $\{\mathbf{x}[n]\}_m \triangleq x(nT + mT_c)$, to digitally capture and compress one instance of the signal according to
\begin{equation}\label{cpx_anlys:eq1}
c_{p}[n] \triangleq \langle x[m], \psi_{p}[m-nM] \rangle.
\end{equation}
For the MF receiver, instead, we assume an oversampling ratio of $2$ to achieve half chip accuracy, i.e., $\Delta \tau = T_c/2$, and downsize the filterbank array. The sequence $\mathbf{x}[n]$ is partitioned into $2$ sub-sequences $\{\mathbf{x}_1[n],\mathbf{x}_2[n]\}$, of $M$ samples each, whose $m$th element is $\{\mathbf{x}_i[n]\}_{m} \triangleq x(nT + mT_c + (i-1)T_c/2)$, $i=1,2$. A typical filter model would process the stream of $2M$ samples sequentially, however to emulate the block processing nature of the CS receiver and avoid CPU cycles that would further delay the execution of the algorithm, we let the $2$ sub-sequences be processed concurrently.

All the arithmetic operations, starting from $\mathbf{x}[n]$, necessary to detect the $|\mathcal{I}|R$ vector elements are recorded and listed in Table \ref{tab1:compx}. The table outlines both single and multiple (MMV) modes for both the MF and CS schemes, and a breakdown of the OMP recovery algorithm adopted by the CS receiver. This popular algorithm seeks the $\mathcal{S}$ (with $|\mathcal{S}|=|\mathcal{I}|R$) non-zero elements of the sparse vector $\bdsb{y}[n]$ by sequentially choosing dictionary elements that better correlate with the observations $\mathbf{c}[n]$. At every iteration the current estimate is subtracted from the observation vector ({\bf OMP.1}) and the residual projected onto the dictionary elements ({\bf OMP.2}). Then, the dictionary element linked to the largest coefficient ({\bf OMP.3}) is retained and removed from the dictionary. The updated set of coefficients is obtained by projecting $\mathbf{c}[n]$ onto the subspace formed by the set of atoms that were removed from the dictionary ({\bf OMP.4}). The algorithm stops when either a maximum number of iterations is reached or when the norm of the residual falls beyond a predefined threshold ({\bf OMP.5}).

\begin{table}
\centering
\begin{tabular}{l*{6}{l}r}
\toprule
\toprule
\textbf{CS Receiver}  & \textbf{Complexity} & \textbf{Remarks} \\
\midrule
Digital compression $c_p[n]$       & $\mathcal{O}(nMP)$  &   Eq. \eqref{cpx_anlys:eq1}     \\
Covariance $\mathbf{R}_{\mathbf{c}\mathbf{c}}$ (optional$^3$)           &  $\mathcal{O}(nP^2)$    &  MMV mode       \\
SVD of $\mathbf{R}_{\mathbf{c}\mathbf{c}}$ (optional$^3$)      & $\mathcal{O}(n^2P)$     &  $n \leq P$ \citep{ConGol90}, MMV mode  \\
Residual update    & $\mathcal{O}(n|\mathcal{S}|^2)$  &  \text{(OMP.1)}   \\
Inner products  & $\mathcal{O}(nPI|\mathcal{K}||\mathcal{Q}||\mathcal{S}|)$    &   $\text{(OMP.2)}$ \\
Maximum projection   & $ \mathcal{O}\big(|\mathcal{S}|\log(I |\mathcal{K}| |\mathcal{Q}|) \big) $     &  $\text{(OMP.3)}$     \\
Least-Squares projection  & $ \mathcal{O}(|\mathcal{S}|^3)  $   &   $ \text{(OMP.4)}$  \\
Stopping criterion   & $ \mathcal{O}(nP |\mathcal{S}|)$     &   $ \text{(OMP.5)} $  \\
\midrule
\midrule
\textbf{MF Receiver}              & \textbf{Complexity} & \textbf{Remarks} \\
\midrule
Correlations $z_{i,k,q}[n]$      &  $\mathcal{O}(nMI|\mathcal{K}||\mathcal{Q}|)$  &  Eq. \eqref{z}   \\
Path selection   &  $\mathcal{O}\big( nIR\ \log(|\mathcal{K}||\mathcal{Q}|) \big)$  &  \\
Accumulation      & $\mathcal{O}(nI|\mathcal{K}||\mathcal{Q}|)$ & MMV mode  \\
\bottomrule
\bottomrule
\end{tabular}
\caption{Complexity breakdown for the proposed CS and traditional MF acquisition using $n$ sets of measurements}\label{tab1:compx}
\end{table}
Path selection refers to identifying the support of a certain vector for pinpointing the active components (delay-Doppler pairs). For both CS and MF, it is tightly coupled to the sorting algorithm being implemented and therefore, we only list its average computational complexity rather than the number of comparators.

When the representation of $\bdsb{y}[n]$ is sufficiently sparse, i.e.  for GPS applications $|\mathcal{S}| \ll I|\mathcal{K}||\mathcal{Q}|$, the number of operations needed to digitally compress $\mathbf{x}[n]$ into $\mathbf{c}[n]$ and to project the residual of each OMP iteration onto the dictionary ({\bf OMP.2}) dominate the overall complexity of the CS receiver, leading to an order of $\mathcal{O}\big(nP(M + I|\mathcal{K}||\mathcal{Q}||\mathcal{S}|)\big)$. On the other hand, the number of operations for the MF are mainly determined by the number of additions to compute the correlations, leading to $\mathcal{O}(nMI|\mathcal{K}||\mathcal{Q}|)$. For hardware implementation this is attractive since the filterbank processing does not require complex multiplications. However, a similar saving can be added to the CS receiver by appropriately designing $\mathbf{B}$ such that $\{\psi_p[m]\}$ also has $\pm 1$ elements.

\begin{figure}[t]
\centering
\includegraphics[width=3in,height=2.5in]{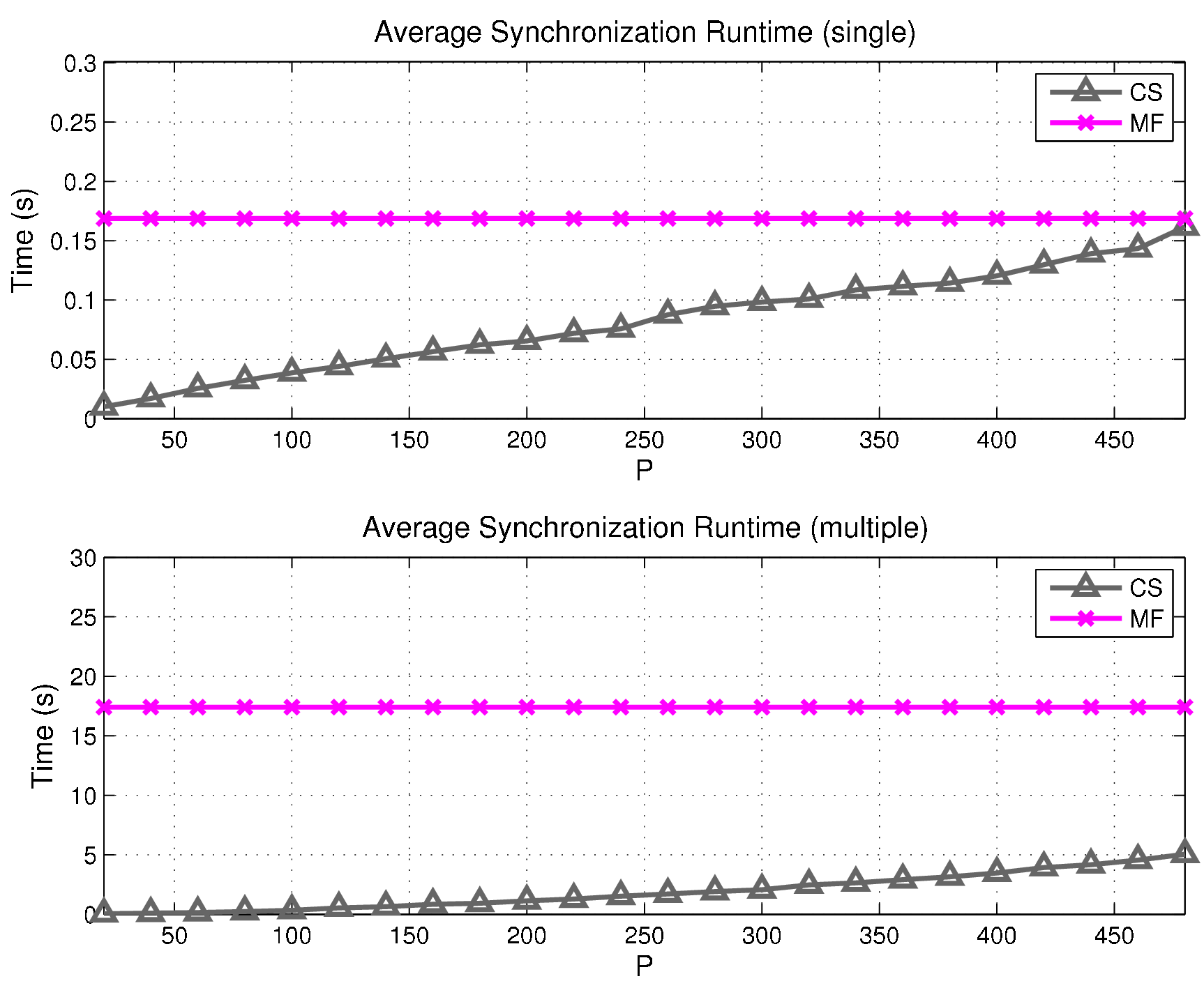}
\caption{\small Average runtime to evaluate $\{\hat{\mathcal{I}},\hat{\bdsb{q}},\hat{\bdsb{k}}\}$ from a received observation vector $\bdsb{y}[n]$ for the compressive scheme, with $n=1$ (above) and $n=50$ (below) as a function of  $P=\{20,40,\ldots,480\}$, and compared against the MF receiver. Each curve was run separately on a 64-bit i7 920 CPU running at 2.67 GHz.}
\label{fig5:smm_offtg_rti}
\end{figure}

The comparison between the dominant terms results in a CS to MF complexity ratio $\big({P}/{I|\mathcal{K}||\mathcal{Q}|} + {P|\mathcal{S}|}/{M}\big)$ that favors the former and emphasizes the complexity savings. In fact, one should in fact expect $P\mathcal{|S|} \ll M$ and $P \ll I|\mathcal{K}||\mathcal{Q}|$, which shows that the CS gains by removing its dependency on the length of the C/A sequence.  This trend is also highlighted in Fig. \ref{fig5:smm_offtg_rti} by the average CPU time spent while executing the steps described in Table \ref{tab1:compx}.

When $n>1$ the ratio remains unchanged since all the additional steps (Table \ref{tab1:compx}) for the ReMBo technique require marginal increase of operations. When compared to $n=1$, the MF spends more CPU time to accumulate the correlation outputs whereas the CS receiver experiences a reverse trend. The additional effort\footnote{Note that in practice, the covariance and SVD computation can be optional by directly choosing a set of measurements $\{\mathbf{c}[n]\}$ and solve the MMV instead.} spent to evaluate $\mathbf{R}_{\mathbf{c}\mathbf{c}}$ is compensated by less operations within the OMP algorithm, and results in a gain in efficiency as highlighted in Table \ref{tab1:compx}.

In general, knowing a priori the order $|\mathcal{S}|$ the CS receiver has an advantage over the MF, which is true and practical in GPS sytems because the order of number of active satellites in the field of view is actually known. However, the MF approach always explores and ranks all the $|\mathcal{K}||\mathcal{Q}|$ dimensions for every satellite before selecting $|\mathcal{I}|R$.

%%%%%%%%%%%
\section{Conclusions}
We proposed a compressive multichannel acquisition scheme for GPS receivers. The reduction is achieved by choosing randomized linear combinations of all the MFs, which leads to great savings in practice. As shown in the analysis and numerical results, our scheme can efficiently recover the unknown delay-Doppler pairs using significantly fewer correlators than those needed in a standard GPS receiver. Regardless of the sparse recovery algorithm, the acquisition performance improves gracefully with the increase of acquisition channels and the number of observations. Therefore, although the proposed scheme has a performance loss in terms of RMSE and success rate compared to the standard GPS scheme, it provides a design tool to trade-off complexity and performance gracefully that can be useful to scale down the cost and energy consumption of GPS chips.

\newpage
\appendix
\section{Proof of Theorem 1}
In this proof, we prove the structure of the matrices $\mathbf{M}_{\phi\phi}(\omega,\bdsb{k},\bdsb{q})$ and $\mathbf{M}_{\phi\phi}(\omega,\mathcal{K},\mathcal{Q})$, which will lead to the results of the proposed theorem. 

Let $\phi_{i,k,q}(-t)=\phi_i(t-qT_c)e^{\mathrm{i}k\Delta\omega t}$. Denote by $\Phi_{i,k,q}(\omega)=\Phi_i(\omega-k\Delta\omega)e^{-\mathrm{i}(\omega-k\Delta\omega) qT_c}$ the Fourier transform of $\phi_{i,k,q}(-t)$. Using $\Phi_i(\omega)=G(\omega)\sum_{m=0}^{M-1}s_i[m]e^{-\mathrm{i}m\omega T_c}$ together with the spectrum $G(\omega)=\left[1+\epsilon_g(\omega)\right]\mathrm{rect}_{2\pi L/T_c}(\omega)$ and ignoring higher order perturbations $\mathcal{O}(|\epsilon_g(\omega)|^2)$, we can write the $[(i',k,q),(i,r)]$th entry of the matrix $\mathbf{M}_{\phi\phi}(\omega,\bdsb{k},\bdsb{q})$ of \eqref{M_GPS} over $\omega\in[-\pi/T,\pi/T]$ as
\begin{align}\label{gen_M_phiphi}
	 \left[\mathbf{M}_{\phi\phi}(\omega,\bdsb{k},\bdsb{q})\right]_{(i',k,q),(i,r)}
	 & = \frac{1}{T}e^{\mathrm{i}\omega(q-q_{i,r})T_c}e^{-\mathrm{i}(kq-k_{i,r}q_{i,r})\Delta\omega T_c}\sum_{\ell=0}^{LM-1} e^{-\mathrm{i}\frac{2\pi\ell}{T}(q-q_{i,r})T_c}\\
	 &\times\sum_{m'=0}^{M-1}\sum_{m=0}^{M-1}s_{i'}^\ast[m']s_i[m]e^{\mathrm{i}\omega(m'-m)T_c} e^{-\mathrm{i}\frac{2\pi\ell}{T}(m'-m)}e^{-\mathrm{i}\Delta\omega T_c(m'k-mk_{i,r})} + \mathcal{O}(\epsilon_g(\omega)).\nonumber
\end{align}
With a change of variable $u=m-m'$, we can re-write the double summations over $m$ and $m'$ as 
\begin{align}
	&\sum_{m'=0}^{M-1}\sum_{m=0}^{M-1}s_{i'}^\ast[m']s_i[m]e^{\mathrm{i}\omega(m'-m)T_c} e^{-\mathrm{i}\frac{2\pi\ell}{T}(m'-m)}e^{-\mathrm{i}\Delta\omega T_c(m'k-mk_{i,r})}\nonumber\\
	=&
	\sum_{u=-M+1}^{M-1} M\cdot\underbrace{\frac{1}{M}\sum_{m=0}^{M-1}s_{i'}^\ast[m-u]s_i[m]e^{-\mathrm{i}\Delta\omega T_c (k-k_{i,r})m}}_{\triangleq R_{i'i}[u, k-k_{i,r}]}e^{-\mathrm{i}u\left(\omega-k\Delta\omega-\frac{2\pi\ell}{T}\right)T_c},
\end{align}
where $R_{i'i}[u, k-k_{i,r}]$ is a pseudo-correlation between the sequence $\{s_{i'}[m]\}$ and $\{s_i[m]\}$ being perturbed by phase-shifts determined by the mismatch of the Doppler shift $e^{-\mathrm{i}\Delta\omega T_c (k-k_{i,r})m}$. Based on the results in \citep{Ivana} and and taking into account that $|s_i[m]|=1$, it can be shown that
\begin{align}
	R_{i'i}[u, k-k_{i,r}] = 
	\begin{cases}
		\displaystyle\frac{1}{M}\sum_{m=0}^{M-1}s_{i'}^\ast[m-u]s_i[m], & k=k_{i,r}\\
		\displaystyle\frac{1}{M}\sum_{m=0}^{M-1}e^{-\mathrm{i}\Delta\omega T_c (k-k_{i,r})m}, & u=0\\
		\mathcal{O}(1/M), & u\neq 0, k \neq k_{i,r},
	\end{cases}
\end{align}
where $\mathcal{O}(1/M)$ is some small perturbation. In particular, it is desirable for the pseudo-correlation $R_{i'i}[u, k-k_{i,r}]$ to decay rapidly over $k$ such that dominant values only appear when there is a frequency component $k\Delta\omega=k_{i,r}\Delta\omega$. By choosing $\Delta\omega = 2\pi j/T$, $j\in\mathbb{Z}^{+}$, we have
\begin{align}
	\frac{1}{M}\sum_{m=0}^{M-1}e^{-\mathrm{i}\Delta\omega T_c (k-k_{i,r})m} = 
	\frac{1}{M}\sum_{m=0}^{M-1}e^{-\mathrm{i} \frac{2\pi j m}{M} (k-k_{i,r})} = \delta[k-k_{i,r}],
\end{align}
which results in a simplified expression of the pseudo-correlation as follows
\begin{align*}
	R_{i'i}[u, k-k_{i,r}] & =
	R_{i'i}[u]\delta[k - k_{i,r}] + \mathcal{O}(1/M),
\end{align*}
where $R_{i'i}[u]$ is the C/A code cross-correlation in \eqref{R_ii}. Therefore, the matrix entry contains significant values only if $k=k_{i,r}$,
\begin{align*}
	\left[\mathbf{M}_{\phi\phi}(\omega,\bdsb{k},\bdsb{q})\right]_{(i,k_{i,r},q),(i,r)} & = \frac{M}{T}e^{\mathrm{i}\omega(q-q_{i,r})T_c}e^{-\mathrm{i}k_{i,r}(q-q_{i,r})\Delta\omega T_c}\sum_{\ell=0}^{LM-1}e^{-\mathrm{i}\frac{2\pi\ell}{T}(q-q_{i,r})T_c}\\
	&~~\times \underbrace{\sum_{u=-M+1}^{M-1}R_{i'i}[u]e^{-\mathrm{i}u\left(\omega-k\Delta\omega -\frac{2\pi\ell}{T}\right)T_c}}_{\triangleq S_{i'i}\left(e^{\mathrm{i}\left(\omega-k\Delta\omega-\frac{2\pi\ell}{T}\right)T_c}\right)} + \mathcal{O}(\epsilon_g(\omega)) + \mathcal{O}(1).
\end{align*}
Furthermore, using the spectrum $S_{i'i}\left(e^{\mathrm{i}\omega T_c}\right)=\delta[i'-i]+\epsilon_{i',i}(\omega)$ and ignoring higher order perturbations $\mathcal{O}(|\epsilon_{i',i}(\omega)|^2)$, the non-zero entries of the matrix $\mathbf{M}_{\phi\phi}(\omega,\bdsb{k},\bdsb{q})$ are explicitly written as
\begin{align*}
	\left[\mathbf{M}_{\phi\phi}(\omega,\bdsb{k},\bdsb{q})\right]_{(i',k_{i,r},q),(i,r)}
	& = \frac{M}{T}e^{\mathrm{i}\omega(q-q_{i,r})T_c}e^{-\mathrm{i}k_{i,r}(q-q_{i,r})\Delta\omega T_c}\sum_{\ell=0}^{LM-1} e^{-\mathrm{i}\frac{2\pi\ell}{T}(q-q_{i,r})T_c} \\
	&~~~+ \mathcal{O}\left(\epsilon_g(\omega)\right)+ \mathcal{O}\left(\epsilon_{i',i}(\omega)\right) + \mathcal{O}(1).
\end{align*}
With $T=MT_c$, we use the property
\begin{align}\label{property}
	\frac{1}{T}\sum_{\ell=0}^{LM-1}e^{-\mathrm{i}\frac{2\pi \ell}{T}(q-q_{i,r})T_c}=L\delta[q-q_{i,r}]
\end{align}
to further express the non-zero entries of $\mathbf{M}_{\phi\phi}(\omega,\bdsb{k},\bdsb{q})$ at $i'=i$, $k=k_{i,r}$ and $q=q_{i,r}$
\begin{align}
	\left[\mathbf{M}_{\phi\phi}(\omega,\bdsb{k},\bdsb{q})\right]_{(i,k_{i,r},q_{i,r}),(i,r)} = LM + \mathcal{O}(\epsilon_g(\omega)) + \mathcal{O}\left(\epsilon_{i,i}(\omega)\right) + \mathcal{O}(1),\quad \omega\in[-\pi/T,\pi/T].
\end{align}
On the other hand, the matrix $\mathbf{M}_{\phi\phi}(\omega,\mathcal{K},\mathcal{Q})$ can be expressed element-wise as
\begin{align*}
	\left[\mathbf{M}_{\phi\phi}(\omega,\mathcal{K},\mathcal{Q})\right]_{(i',k',q'),(i,k,q)}  &= \frac{1}{T}\sum_{\ell=0}^{LM-1}
	   \Phi_{i'}^\ast\left(\omega-k'\Delta\omega-\frac{2\pi \ell}{T}\right)e^{\mathrm{i}\left(\omega-k'\Delta\omega-\frac{2\pi \ell}{T}\right)q'T_c}\\
	   &~~~~~~~\times \Phi_i\left(\omega-k\Delta\omega-\frac{2\pi \ell}{T}\right) e^{-\mathrm{i}\left(\omega-k\Delta\omega-\frac{2\pi \ell}{T}\right)qT_c}.
\end{align*}
Similarly, the expression is significant only when $k=k'$
\begin{align*}
	\left[\mathbf{M}_{\phi\phi}(\omega,\mathcal{K},\mathcal{Q})\right]_{(i',k,q'),(i,k,q)} & =  \frac{1}{T}e^{\mathrm{i} \omega(q'-q)T_c}e^{\mathrm{i}k(q-q')\Delta\omega T_c} \sum_{\ell=0}^{LM-1}e^{\mathrm{i} \frac{2\pi \ell}{T} (q-q')T_c} + \mathcal{O}(\epsilon_g(\omega)) + \mathcal{O}(\epsilon_{i',i}(\omega)) + \mathcal{O}(1).
\end{align*}
When $q=q'$, according to \eqref{property}, we have
\begin{align*}
	\left[\mathbf{M}_{\phi\phi}(\omega,\mathcal{K},\mathcal{Q})\right]_{(i',k,q),(i,k,q)} & = LM + \mathcal{O}(\epsilon_g(\omega)) + \mathcal{O}(\epsilon_{i',i}(\omega)) + \mathcal{O}(1).
\end{align*}
Since the error functions satisfy $\|\epsilon_g(\omega)\|\ll 1$ and $\|\epsilon_{i',i}(\omega)\|\ll 1$, the results in Theorem 1 follow.

\end{document}